\documentclass[journal]{IEEEtran}

\usepackage{mathrsfs}
\usepackage{dsfont}
\usepackage{amsfonts}
\usepackage{changepage}
\usepackage{bm}
\usepackage{booktabs}
\usepackage{multirow}
\usepackage{makecell}

\usepackage{textcomp,booktabs,threeparttable, amssymb}
\usepackage[usenames,dvipsnames]{color}
\usepackage{colortbl}
\definecolor{mygray}{gray}{.9}
\definecolor{mypink}{rgb}{.99,.91,.95}
\definecolor{mycyan}{cmyk}{.3,0,0,0}

\usepackage[pdftex]{graphicx}
\graphicspath{{../pdf/}{../jpeg/}}
\DeclareGraphicsExtensions{.pdf,.jpeg,.png}

\usepackage[cmex10]{amsmath}
\usepackage{array}
\usepackage{mdwmath}
\usepackage{eqparbox}
\usepackage{url}
\usepackage{hyperref}
\usepackage[hyphenbreaks]{breakurl}

\begin{document}
\bstctlcite{IEEEexample:BSTcontrol}
    \title{Nonparametric Multivariate Probability Density Forecast in Smart Grids With Deep Learning}
  \author{Zichao Meng,~\IEEEmembership{Student Member,~IEEE,}
      Ye Guo,~\IEEEmembership{Senior Member,~IEEE,}\\
      Wenjun Tang,~\IEEEmembership{Member,~IEEE,}
      and~Hongbin Sun,~\IEEEmembership{Fellow,~IEEE,}

  }

\markboth{IEEE TRANSACTIONS ON POWER SYSTEMS
}{Roberg \MakeLowercase{\textit{et al.}}: Nonparametric Multivariate Probability Density Forecast in Smart Grids With Deep Learning}

\maketitle

\setcounter{equation}{27}

\section*{Appendix}
\appendices

\section{Failure of DAN-NFN in Modeling the Joint Probability Distribution}
\label{AA} 

We change the DAN part in DAN-NFN from a SISO positive-weighted NN to a MISO one to reformulate DAN-NFN in a multivariate setting preparing for the multivariate density forecast. For conciseness, denote $\mathbf{y}=[\text{y}_1\text{, }\text{y}_2\text{, }\cdots\text{, }\text{y}_D]$ as the input to DAN, which represents a vector with $D$ variables. We construct the mapping of the reformulated DAN (a MISO positive-weighted NN) as
\begin{equation}
\begin{array}{l}
\mathbf{\Psi}(\mathbf{y}\text{; }\mathbf{w}^+\text{, }\mathbf{b})=\\
z\{\mathbf{w}_{K+1}^+\cdots z[\mathbf{w}_2^+\cdot z(\mathbf{w}_1^+\cdot\mathbf{y}+\mathbf{b}_1)+\mathbf{b}_2]\cdots+\mathbf{b}_{K+1}\}.
\end{array}
\end{equation}
where $K$ is the number of hidden layers. $z$ is the activation function (sigmoid, tanh, linear, or ReLU), and can be different in different layers. Tensors $\mathbf{w}^+$ and $\mathbf{b}$ represent all the weights and biases, respectively, which are determined by the outputs of NFN. Tensors $\mathbf{w}_k^+\in\mathbf{w}^+$ and $\mathbf{b}_k\in\mathbf{b}$, $k\in[1, K+1]$, are weights and biases in $k$th layer, respectively.

For $k=1\text{, }\cdots\text{, }K$, define the input-output mapping of layer $k$ in (28) as
\begin{equation}
\begin{aligned}
&\mathbf{Y}_k=z(\mathbf{w}_k^+\cdot\mathbf{Y}_{k-1}+\mathbf{b}_k)\\ &=
\begin{bmatrix}
   z(\mathbf{w}_{k\text{, }1}^+\cdot\mathbf{Y}_{k-1\text{, }1}+\mathbf{b}_{k\text{, }1})\\
  \vdots\\
  z(\mathbf{w}_{k\text{, }l}^+\cdot\mathbf{Y}_{k-1\text{, }l}+\mathbf{b}_{k\text{, }l})\\
   \vdots
\end{bmatrix}
=
\begin{bmatrix}
   z_{k\text{, }1}\\
  \vdots\\
  z_{k\text{, }l}\\
   \vdots
\end{bmatrix}\text{,}
\end{aligned}
\end{equation}
\begin{equation}
\mathbf{Y}_{K+1}=z(\mathbf{w}_{K+1}^+\cdot\mathbf{Y}_K+\mathbf{b}_{K+1})\text{,}
\end{equation}
where $\mathbf{Y}_0=\mathbf{y}$. Tensors $\mathbf{w}_{k\text{, }l}^+$ and $\mathbf{b}_{k\text{, }l}$ represent the $l$th row in $\mathbf{w}_{K+1}^+$ and $\mathbf{b}_{K+1}$, respectively. $z_{k\text{, }l}$ is short for $z(\mathbf{w}_{k\text{, }l}^+\cdot\mathbf{Y}_{k-1\text{, }l}+\mathbf{b}_{k\text{, }l})$. Note that $\mathbf{w}_{K+1}^+$ and $\mathbf{b}_{K+1}$ in the output layer are vector and scalar, respectively. Then, for $k$=1, $\cdots$, $K$:
\begin{equation}
\frac{d\mathbf{Y}_k}{d\mathbf{Y}_{k-1}}=
\begin{bmatrix}
   z_{k\text{, }1}^\prime(\mathbf{w}_{k\text{, }1}^+)^\mathrm{T}&\cdots&z_{k\text{, }l}^\prime(\mathbf{w}_{k\text{, }l}^+)^\mathrm{T}&\cdots
\end{bmatrix}\text{,}
\end{equation}
\begin{equation}
\frac{d\mathbf{Y}_{K+1}}{d\mathbf{Y}_K}=
   z_{K+1}^\prime(\mathbf{w}_{K+1}^+\cdot\mathbf{Y}_K+\mathbf{b}_{K+1})(\mathbf{w}_{K+1}^+)^\mathrm{T}.
\end{equation}

Thus,
\begin{equation}
\frac{d\mathbf{\Psi}(\mathbf{y}\text{; }\mathbf{w}^+\text{, }\mathbf{b})}{d\mathbf{y}}=\prod_{k=1}^{K+1}\frac{d\mathbf{Y}_k}{d\mathbf{Y}_{k-1}}.
\end{equation}

The first derivatives of different activation function $z$ are
\begin{equation}
z^\prime=\left\{
\begin{aligned}
&\sigma\cdot(1-\sigma)\text{, }&&\text{if $z$ is sigmoid}\\
&2\sigma\cdot(1-\sigma)\text{, }&&\text{if $z$ is tanh}\\
&1\text{, }&&\text{if $z$ is linear}\\
&0\text{ or }1\text{, }&&\text{if $z$ is ReLU}\\
\end{aligned}
\right.
\end{equation}
where $\sigma$ denotes the sigmoid function. Since $\sigma\in(0, 1)$, $z^\prime$ is nonnegative for all kinds of activation functions considered in this paper. Considering that every entry in $\mathbf{w}^+$ is positive and combining (31)-(34), one can verify that $d\mathbf{\Psi}(\mathbf{y}\text{; }\mathbf{w}^+\text{, }\mathbf{b})/d\mathbf{y}$ is a $d$-dimensional vector, and any entry in it is positive. Therefore, $\mathbf{\Psi}(\mathbf{y}\text{; }\mathbf{w}^+\text{, }\mathbf{b})$ is multivariate monotonically nondecreasing.

Now, we illustrate that why such multivariate monotonically nondecreasing property can not be extended to the higher-order-derivative form that meets condition (ii). For conciseness, an example is taken when there are two layers in the positive-weighted NN, which can be denoted as
\begin{equation}
\mathbf{\Gamma}=z[\mathbf{w}_2^+\cdot z(\mathbf{w}_1^+\cdot\mathbf{y}+\mathbf{b}_1)+\mathbf{b}_2].
\end{equation}

Defining $y_p$ and $y_q$ are arbitrary two entries in $\mathbf{y}$, based on the analyses about (31)-(34), the second-order partial derivative of $\mathbf{\Gamma}$ with respect to them can be derived as
\begin{equation}
\begin{small}
\begin{aligned}
\frac{\partial^2\mathbf{\Gamma}}{\partial y_p\partial y_q}=&
\begin{bmatrix}
   z_{1\text{, }1}^{\prime\prime}\cdot\mathbf{w}_{1\text{, }1p}^+\cdot\mathbf{w}_{1\text{, }1q}^+&\cdots&z_{1\text{, }l}^{\prime\prime}\cdot\mathbf{w}_{1\text{, }lp}^+\cdot\mathbf{w}_{1\text{, }lq}^+&\cdots
\end{bmatrix}\\
&\cdot z^\prime(\mathbf{w}_2^{+}\cdot\mathbf{Y}_2+\mathbf{b}_2)\cdot(\mathbf{w}_2^+)^{\mathrm{T}}\\
&+\begin{bmatrix}
   z_{1\text{, }1}^{\prime}\cdot\mathbf{w}_{1\text{, }1p}^+&\cdots&z_{1\text{, }l}^{\prime}\cdot\mathbf{w}_{1\text{, }lp}^+&\cdots
\end{bmatrix}\cdot(\mathbf{w}_2^+)^\mathrm{T}\\
&\cdot\begin{bmatrix}
   z_{1\text{, }1}^{\prime}\cdot\mathbf{w}_{1\text{, }1q}^+&\cdots&z_{1\text{, }l}^{\prime}\cdot\mathbf{w}_{1\text{, }lq}^+&\cdots
\end{bmatrix}\\
&\cdot z^{\prime\prime}(\mathbf{w}_2^+\cdot\mathbf{Y}_2+\mathbf{b}_2)
\cdot(\mathbf{w}_2^+)^\mathrm{T}
\text{,}
\end{aligned}
\end{small}
\end{equation}
where $\mathbf{w}_{1\text{, }lp}\text{ }(\mathbf{w}_{1\text{, }lq})$ is the element at $l$th row and $p$th ($q$th) column in $\mathbf{w}_1^+$. The second derivatives of different activation
function $z$ are
\begin{equation}
z^{\prime\prime}=\left\{
\begin{aligned}
&\sigma\cdot(1-\sigma)\cdot(1-2\sigma)\text{, }&&\text{if $z$ is sigmoid}\\
&2\sigma\cdot(1-\sigma)\cdot(1-2\sigma)\text{, }&&\text{if $z$ is tanh}\\
&0\text{. }&&\text{if $z$ is linear or ReLU}
\end{aligned}
\right.
\end{equation}

It shows that the nonnegative property does not always hold for $z^{\prime\prime}$. For sigmoid or tanh activation function, $z^{\prime\prime}$ will be negative if the intermittent computing result is greater than zero when doing forward or backward propagation in the network, which is observed very commonly. Although $z^{\prime\prime}$ can be nonnegative all the time for ReLU or linear activation function, the NN still could learn nothing because the gradients are always zero. Therefore, combining (36) and (37), condition (ii) can not be guaranteed.

A simple idea is to find a very special activation function so that
\begin{equation}
z^\prime\geq0\text{, }z^{\prime\prime}\geq0\text{, }\cdots\text{, }z^{(D)}\geq0\text{, }
\end{equation}
which ensures condition (ii). One activation function satisfying (38) is the exponential function $(e^x)$. However, it is rarely used in NNs as exponential function suffers from vanishing/exploding gradient problems easily.

Based on the analyses above, simply replacing DAN in the original DAN-NFN framework as a MISO positive-weighted NN can not be used to represent the joint probability distribution.

\section{Proof of the Nonnegativity of Equation (12)}
\label{AB} 
The proof is derived by induction on the value of $D$, i.e., the number of random variables in forecasting targets. For conciseness, denoting $\boldsymbol{y}=[y^1\text{, }y^2\text{, }\cdots\text{, }y^D]$ as the input to JDAN, the output of $d$th parallel unit after the corresponding normalization layer can be reformulated from (9) to
\begin{equation}
\overline{\mathbf{\Psi}}^d=\frac{\mathbf{\Psi}^d(y^{d}\text{; }\mathbf{W}^{d+}\text{, }\mathbf{B}^d)-\mathbf{\Psi}^d(L_d\text{; }\mathbf{W}^{d+}\text{, }\mathbf{B}^d)}{\mathbf{\Psi}^d(U_d\text{; }\mathbf{W}^{d+}\text{, }\mathbf{B}^d)-\mathbf{\Psi}^d(L_d\text{; }\mathbf{W}^{d+}\text{, }\mathbf{B}^d)}\text{,}
\end{equation}

In a base case of $D=2$ and $\boldsymbol{y}=[y^1\text{, }y^2]$, the input-output mapping of JDAN, denoted as $\mathbf{\Psi}_\mathcal{J}^{1,2}$, can be represented as
\begin{equation}
    \mathbf{\Psi}_\mathcal{J}^{1,2}=\overline{\mathbf{\Psi}}^1\cdot\overline{\mathbf{\Psi}}^2\cdot[\mathbf{C}_{12}\cdot(1-\overline{\mathbf{\Psi}}^1)\cdot(1-\overline{\mathbf{\Psi}}^2)+1].
  \end{equation}

Then, the second-order partial derivative of $\mathbf{\Psi}_\mathcal{J}^{1,2}$ with respect to [$y^1$, $y^2$] can be derived as
\begin{equation}
  \frac{\partial^{2}(\mathbf{\Psi}_\mathcal{J}^{1,2})}{\partial y^{1}\partial y^{2}}=\frac{\partial\overline{\mathbf{\Psi}}^1}{\partial y^{1}}\cdot\frac{\partial\overline{\mathbf{\Psi}}^2}{\partial y^{2}}\cdot\left[\mathbf{C}_{12}\cdot(1-2\overline{\mathbf{\Psi}}^1)\cdot(1-2\overline{\mathbf{\Psi}}^2)+1\right].
  \end{equation}

Since $\overline{\mathbf{\Psi}}^1$, $\overline{\mathbf{\Psi}}^2\in[0,1]$, $\mathbf{C}_{12}\in(-1,1)$, and ${\partial\overline{\mathbf{\Psi}}^1}/{\partial y^{1}}$, ${\partial\overline{\mathbf{\Psi}}^2}/{\partial y^{2}}\geq0$, we have $\frac{\partial^{2}(\mathbf{\Psi}_\mathcal{J}^{1,2})}{\partial y^{1}\partial y^{2}}\geq0$. Note that this nonnegative property holds for any two random variables.

Next, for $D=m$, the mapping of JDAN, denoted as $\mathbf{\Psi}_\mathcal{J}^{m}$, is given as
\begin{equation}
  \begin{aligned}
    \mathbf{\Psi}_\mathcal{J}^{m}&=\prod_{d=1}^m\overline{\mathbf{\Psi}}^d\cdot\sum_{i>d}^{m}\sum_{d=1}^{m-1}[\mathbf{C}_{di}\cdot(1-\overline{\mathbf{\Psi}}^d)\cdot(1-\overline{\mathbf{\Psi}}^i)+1]\\
  &=g(\overline{\mathbf{\Psi}})\cdot h(\overline{\mathbf{\Psi}})\text{,}
  \end{aligned}
  \end{equation}
where $g(\overline{\mathbf{\Psi}})=\prod_{d=1}^m\overline{\mathbf{\Psi}}^d$, $h(\overline{\mathbf{\Psi}})=\sum_{i>d}^{m}\sum_{d=1}^{m-1}[\mathbf{C}_{di}\cdot(1-\overline{\mathbf{\Psi}}^d)\cdot(1-\overline{\mathbf{\Psi}}^i)+1]$, and we omit the constant coefficient $\frac{1}{\binom{m}{2}}$ for brevity.

Assuming that the nonnegative property holds for $D=m$, $m>2$, we have 
\begin{equation}
\frac{\partial^{m}(\mathbf{\Psi}_\mathcal{J}^m)}{\partial y^{1}\cdots\partial y^{m}}\geq0
\end{equation}

Now, for $D=m+1$, the mapping of JDAN can be inferred as 
\begin{equation}
  \begin{aligned}
    &\mathbf{\Psi}_\mathcal{J}^{m+1}=g(\overline{\mathbf{\Psi}})\cdot\overline{\mathbf{\Psi}}^{m+1}\cdot \\ 
    &\left\{h(\overline{\mathbf{\Psi}})+\sum_{d=1}^{m}\left[\mathbf{C}_{d(m+1)}\cdot(1-\overline{\mathbf{\Psi}}^d)\cdot(1-\overline{\mathbf{\Psi}}^{m+1})+1\right]\right\}.\\
    &=\mathbf{\Psi}_\mathcal{J}^{m}\cdot\overline{\mathbf{\Psi}}^{m+1}+\sum_{d=1}^m\frac{g(\overline{\mathbf{\Psi}})}{\overline{\mathbf{\Psi}}^d}\cdot\overline{\mathbf{\Psi}}^d\cdot\overline{\mathbf{\Psi}}^{m+1}\cdot\\
    &\left[\mathbf{C}_{d(m+1)}\cdot(1-\overline{\mathbf{\Psi}}^d)\cdot(1-\overline{\mathbf{\Psi}}^{m+1})+1\right]\\
    &=\mathbf{\Psi}_\mathcal{J}^{m}\cdot\overline{\mathbf{\Psi}}^{m+1}+\sum_{d=1}^m\prod_{\substack{i=1\\i\neq d}}^{m}\overline{\mathbf{\Psi}}^{i}\cdot\mathbf{\Psi}_\mathcal{J}^{d,m+1}
  \end{aligned}
  \end{equation}

The higher-order partial derivative of $\mathbf{\Psi}_\mathcal{J}^{m+1}$ with respect to $m+1$ random variables can be derived as
\begin{equation}
  \begin{aligned}
  &\frac{\partial^{m+1}(\mathbf{\Psi}_\mathcal{J}^{m+1})}{\partial y^{1}\cdots\partial y^{m}\partial y^{m+1}}=\frac{\partial^{m}(\mathbf{\Psi}_\mathcal{J}^m)}{\partial y^{1}\cdots\partial y^{m}}\cdot\frac{\partial \overline{\mathbf{\Psi}}^{m+1}}{\partial y^{m+1}}+\\
  &\sum_{d=1}^m\prod_{\substack{i=1\\i\neq d}}^{m}\frac{\partial \overline{\mathbf{\Psi}}^{i}}{\partial y^{i}}\cdot\frac{\partial^2\mathbf{\Psi}_\mathcal{J}^{d,m+1}}{\partial y^d \partial y^{m+1}}.
  \end{aligned}
  \end{equation}

Combining (41) and (43), we have $\frac{\partial^{m+1}(\mathbf{\Psi}_\mathcal{J}^{m+1})}{\partial y^{1}\cdots\partial y^{m}\partial y^{m+1}}\geq0$, and we can conclude that the nonnegative property still holds for $D=m+1$. 

Therefore, we have proved the nonnegativity of (12) for $D\geq2$.








\ifCLASSOPTIONcaptionsoff
  \newpage
\fi

\vfill



\begin{thebibliography}{00}

\bibitem{IRENA-2020}
IRENA (2020), Renewable Energy Statistics 2020, \emph{The International Renewable Energy Agency,} Abu Dhabi. Accessed: Dec. 21, 2021. [Online]. Available: https://www.irena.org/-/media/Files/IRENA/Agency/Publication/2020/Jul/IRENA\_Renewable\_\\energy\_statistics\_2020.pdf

\bibitem{AEE-2021}
Agora Energiewende and Ember (2021): \emph{The European Power Sector in 2020: Up-to-Date Analysis on the Electricity Transition.} Accessed: Dec. 21, 2021. [Online]. Available: https://static.agora-energiewende.de/fileadmin/Projekte/2021/2020\_01\_EU-Annual-Review\_2020/A-EW\_202\_Report\_European-Power-Sector-2020.pdf

\bibitem{Wang-PSCE-2021}
J. Wang \emph{et al.}, ``Building load forecasting using deep neural network with efficient feature fusion," \emph{Journal of Modern Power Systems and Clean Energy,} vol. 9, no. 1, pp. 160-169, Jan. 2021.

\bibitem{Jahangir-TIE-2021}
H. Jahangir \emph{et al.}, ``Deep learning-based forecasting approach in smart grids with microclustering and bidirectional LSTM network," \emph{IEEE Transactions on Industrial Electronics,} vol. 68, no. 9, pp. 8298-8309, Sept. 2021.

\bibitem{Zhang-TSG-2019}
W. Zhang \emph{et al.}, ``An improved quantile regression neural network for probabilistic load forecasting," \emph{IEEE Transactions on Smart Grid,} vol. 10, no. 4, pp. 4425-4434, Jul. 2019.

\bibitem{Wen-TNLS-2020}
Y. Wen \emph{et al.}, ``Performance evaluation of probabilistic methods based on bootstrap and quantile regression to quantify PV power point forecast uncertainty," \emph{IEEE Transactions on Neural Networks and Learning Systems,} vol. 31, no. 4, pp. 1134-1144, Apr. 2020.

\bibitem{Salinas-online-2017}
M. Afrasiabi \emph{et al.}, ``Deep-based conditional probability density function forecasting of residential loads," \emph{IEEE Transactions on Smart Grid,} vol. 11, no. 4, pp. 3646-3657, Jul. 2020.

\bibitem{Orozco-online-2018}
C. Wan \emph{et al.}, ``An adaptive ensemble data driven approach for nonparametric probabilistic forecasting of electricity load," \emph{IEEE Transactions on Smart Grid,} vol. 12, no. 6, pp. 5396-5408, Nov. 2021.

\bibitem{Zhu-ICDMW-2017}
L. Zhu \emph{et al.}, ``Deep and confident prediction for time series at Uber,'' in \emph{Proc. IEEE International Conference on Data Mining Workshops (ICDMW),} pp. 103-110, Nov. 2017.

\bibitem{Wang-AE-2017}
H. Wang \emph{et al.}, ``Deep learning based ensemble approach for probabilistic wind power forecasting,'' \emph{Applied energy,} vol. 188, pp: 56-70, Feb. 2017.

\bibitem{Kilian-book-2017}
 L. Kilian \emph{et al.} \emph{Structural vector autoregressive analysis.} Cambridge University Press, 2017.

\bibitem{Zhang-AA-2021}
Y. Zhang \emph{et al.}, ``Multivariate probabilistic forecasting and its performance's impacts on long-term dispatch of hydro-wind hybrid systems," \emph{Applied Energy,} vol. 283, no. 7, pp. 1-22, Feb. 2021.

\bibitem{Wang1-TPS-2018}
Z. Wang \emph{et al.}, ``Probabilistic forecast for multiple wind farms based on regular vine copulas," \emph{IEEE Transactions on Power Systems,} vol. 33, no. 1, pp. 578-589, Jan. 2018.

\bibitem{Wang-MPSC-2020}
Z. Wang \emph{et al.}, ``Forecasted scenarios of regional wind farms based on regular vine copulas," \emph{Journal of Modern Power Systems and Clean Energy,} vol. 8, no. 1, pp. 77-85, Jan. 2020.

\bibitem{Golestaneh-ISGT-2017}
F. Golestaneh \emph{et al.}, ``Multivariate prediction intervals for photovoltaic power generation," in \emph{Proc. IEEE Innovative Smart Grid Technologies - Asia (ISGT-Asia),}, Dec. 2017, pp. 1-5.

\bibitem{Toubeau-TPS-2019}
J. Toubeau \emph{et al.}, ``Deep learning-based multivariate probabilistic forecasting for short-term scheduling in power markets," \emph{IEEE Transactions on Power Systems,} vol. 34, no. 2, pp. 1203-1215, Mar. 2019.

\bibitem{Charytoniuk-TPS-1998}
W. Charytoniuk \emph{et al.}, ``Nonparametric regression based short-term load forecasting," \emph{IEEE Transactions on Power Systems,} vol. 13, no. 3, pp. 725-730, Aug. 1998.

\bibitem{Xu-TSE-2020}
X. Xu \emph{et al.}, ``Data-driven risk-averse two-stage optimal stochastic scheduling of energy and reserve with correlated wind power," \emph{IEEE Transactions on Sustainable Energy,} vol. 11, no. 1, pp. 436-447, Jan. 2020.

\bibitem{Little-TIA-2018}
M. L. Little \emph{et al.}, ``Unified probabilistic modeling of wind reserves for demand response and frequency regulation in islanded microgrids," \emph{IEEE Transactions on Industry Applications,} vol. 54, no. 6, pp. 5671-5681, Nov.-Dec. 2018.

\bibitem{Bracale-TPS-2020}
A. Bracale \emph{et al.}, ``Multivariate quantile regression for short-term probabilistic load forecasting," \emph{IEEE Transactions on Power Systems,} vol. 35, no. 1, pp. 628-638, Jan. 2020.

\bibitem{Craiu-AAS-2008}
M. Craiu \emph{et al.}, ``On the choice of parametric families of copulas." \emph{Advances \& Applications in Statistics,} vol. 10, no. 1, pp. 25-40, Nov. 2008.

\bibitem{Crabbe-book-2013}
J. J. Crabbe \emph{et al.} \emph{Handling the curse of dimensionality in multivariate kernel density estimation.} Available from ProQuest Dissertations \& Theses Global, SciTech Premium Collection, 2013.

\bibitem{Strelen-conf-2007}
J. C. Strelen \emph{et al.}, ``Analysis and generation of random vectors with copulas,'' in \emph{Proc. 2007 Winter Simulation Conference,} 2007, pp. 488-496.

\bibitem{Hu-TNLS-2020}
T. Hu \emph{et al.}, ``Distribution-free probability density forecast through deep neural networks,'' \emph{IEEE Transactions on Neural Networks and Learning Systems,} vol. 31, no. 2, pp. 612-625, Feb. 2020.

\bibitem{Montufar-ANIPS-2014}
G. F. Montufar \emph{et al.}, ``On the number of linear regions of deep neural networks,'' in \emph{Proc. Advances in Neural Information Processing Systems,} 2014, pp. 2924-2932.

\bibitem{Daniels-TNLS-2010}
H. Daniels \emph{et al.}, ``Monotone and partially monotone neural networks,'' \emph{IEEE Transactions on Neural Networks,} vol. 21, no. 6, pp. 906-917, Jun. 2010.

\bibitem{He-ANIPS-2014}
K. He \emph{et al.}, ``Deep residual learning for image recognition,'' in \emph{Proc. IEEE Conference on Computer Vision and Pattern Recogintion,} Jun. 2016, pp. 770-778.

\bibitem{Greff-TNLS-2017}
K. Greff \emph{et al.}, ``LSTM: A search space odyssey,'' \emph{IEEE Transactions on Neural Networks and Learning Systems,} vol. 28, no. 10, pp. 2222-2232, Oct. 2017.

\bibitem{Ioffe-ICML-2015}
S. Ioffe \emph{et al.}, ``Batch normalization: Accelerating deep network training by reducing internal covariate shift,'' in \emph{Proc. International Conference on Machine Learning,} 2015, pp. 448-456.

\bibitem{Pinson-WE-2007}
P. Pinson \emph{et al.}, ``Non-parametric probabilistic forecasts of wind power: Required properties and evaluation,'' \emph{Wind Energy,} vol. 10, no. 6, pp. 497-516, Nov./Dec. 2007.

\bibitem{Scheuerer-MWR-2015}
M. Scheuerer \emph{et al.}, ``Variogram-based proper scoring rules for probabilistic forecasts of multivariate quantities,'' \emph{Monthly Weather Review,} vol. 143, no. 4,pp. 1321-1334, Apri. 2015.

\bibitem{West-Texas-Mesonet}
\emph{West Texas Mesonet.} Accessed: Jun. 1, 2021. [Online]. Available: http://meso-file1.tosm.ttu.edu/tech/1-output/mesonet.php

\bibitem{Generation-data}
\emph{Actual Generation and Load Data.} Accessed: Dec. 21, 2021. [Online]. Available: https://www.aemo.com.au/energy-systems/electricity/national-electricity-market-nem/data-nem/market-management-system-mms-data/generation-and-load

\bibitem{Demand-data}
\emph{Aggregated Price and Demand Data.} Accessed: Dec. 21, 2021. [Online]. Available: https://www.aemo.com.au/energy-systems/electricity/national-electricity-market-nem/data-nem/aggregated-data

\bibitem{Genest-HE-2007}
C. Genest \emph{et al.}, ``Everything you always wanted to know about copula modeling but were afraid to ask,'' \emph{Journal of Hydrologic Engineering,} vol. 12, no. 4, pp. 347-368, Jul. 2007.

\bibitem{Moller-CSDA-2008}
J. K. M\o ller \emph{et al.}, ``Time-adaptive quantile regression,'' \emph{Computational Statistics and Data Analysis,} vol. 52, no. 3, pp. 1292-1303, Jan. 2008.

\bibitem{Zhang-RSER-2014}
Y. Zhang \emph{et al.}, ``Review on probabilistic forecasting of wind power generation,'' \emph{Renewable and Sustainable Energy Reviews,} vol 32, pp. 255-270, Apr. 2014.

\bibitem{Wei-QREF-2011}
W. Gregor, ``Are copula-GoF-tests of any practical use? Empirical evidence for stocks, commodities and FX futures,'' \emph{The Quarterly Review of Economics and Finance,}
vol. 51, no. 2, pp. 173-188, May 2011.

\bibitem{Duong-JNS-2010}
T. Duong \emph{et al.}, ``Plug-in bandwidth matrices for bivariate kernel density estimation,'' \emph{Journal of Nonparametric Statistics,} vol. 15, no. 1, pp. 17-30, Oct. 2010. 

\bibitem{Langren-JCGS-2018}
N. Langrené \emph{et al.}, ``Fast and stable multivariate kernel density estimation by fast sum updating,'' \emph{Journal of Computational and Graphical Statistics,} vol. 28, no. 3, pp. 596-608, Nov. 2018.

\bibitem{Taieb-TSG-2016}
S. B. Taieb \emph{et al.}, ``Forecasting Uncertainty in Electricity Smart Meter Data by Boosting Additive Quantile Regression,'' \emph{IEEE Transactions on Smart Grid,} vol. 7, no. 5, pp. 2448-2455, Sept. 2016.

\bibitem{Gamboa-arXiv-2017}
J. Gamboa, ``Deep learning for time-series analysis,'' \emph{arXiv preprint arXiv: 1701.01887,} 2017. 

\bibitem{Kuremoto-Neuro-2014}
T. Kuremoto \emph{et al.}, ``Time series forecasting using a deep belief network with restricted Boltzmann machines,'' \emph{Neurocomputing,} vol. 137, pp. 47-56, Aug. 2014. 

\bibitem{Tian-Neuro-2018}
Y. Tian \emph{et al.}, ``LSTM-based traffic flow prediction with missing data,'' \emph{Neurocomputing,}  vol. 318, pp. 297-305, Nov. 2018.

\bibitem{Mussumeci-SSE-2020}
E. Mussumeci \emph{et al.}, ``Large-scale multivariate forecasting models for Dengue-LSTM versus random forest regression,'' \emph{Spatial and Spatio-temporal Epidemiology,} vol. 35, Nov. 2020.

\bibitem{Shih-ML-2019}
S. Shih \emph{et al.}, ``Temporal pattern attention for multivariate time series forecasting,'' \emph{Machine Learning,} vol. 108, pp. 1421-1441, Jun. 2019.

\bibitem{Du-Neuro-2020}
S. Du \emph{et al.}, ``Multivariate time series forecasting via attention-based encoder-decoder framework,'' \emph{Neurocomputing,} vol. 388, pp. 269-279, May 2020.

\bibitem{Rangapuram-NIPS-2018}
S. Rangapuram \emph{et al.}, ``Deep state space models for time series forecasting,'' \emph{Advances in neural information processing systems,} vol. 31, 2018.

\bibitem{Chen-TSE-2020}
J. Chen \emph{et al.}, ``Learning heterogeneous features jointly: A deep end-to-end framework for multi-step short-term wind power prediction,'' \emph{IEEE Transactions on Sustainable Energy,} vol 11, no.3, pp. 1761-1772, Jul. 2020.


\bibitem{Zhang-TSG-2020}
W. Zhang \emph{et al.}, ``Improving Probabilistic Load Forecasting Using Quantile Regression NN With Skip Connections,'' \emph{IEEE Transactions on Smart Grid,} vol. 11, no. 6, pp. 5442-5450, Nov. 2020.

\bibitem{Salinas-IJF-2020}
D. Salinas \emph{et al.}, ``DeepAR: Probabilistic forecasting with autoregressive recurrent networks,'' \emph{International Journal of Forecasting,} vol. 36, no. 3, pp. 1181-1191, 2020.

\end{thebibliography}
\end{document}